\documentclass[aps,prb,twocolumn,showpacs,superscriptaddress,amsmath,amssymb,amsfonts,floatfix,longbibliography]{revtex4-2}

\usepackage{siunitx} %
\usepackage{braket} %
\usepackage{extarrows} %
\usepackage{bm} %
\usepackage{nicematrix} %

\newcommand{\degree}{^\circ}

\newcommand{\md}{\mathrm{d}}

\usepackage[T1]{fontenc}
\usepackage{graphicx} %
\usepackage{dcolumn} %
\usepackage{listings} %
\usepackage{xcolor}
\usepackage{multirow}
\usepackage[
    colorlinks,
    linkcolor=blue,
    filecolor=magenta,
    urlcolor=cyan,
    citecolor=blue
]{hyperref}%

\begin{document}

\title{Ferrichiral skyrmions with sublattice-resolved chirality in extended Kitaev  model in triangular lattice} 
\author{Bogeng Wen}
    \affiliation{Department of Physics, University of Toronto, Toronto, Ontario M5S 1A7, Canada}
\author{Jiefu Cen}
    \affiliation{Department of Physics, University of Toronto, Toronto, Ontario M5S 1A7, Canada}
\author{Hae-Young Kee}
    \email{hy.kee@utoronto.ca}
    \affiliation{Department of Physics, University of Toronto, Toronto, Ontario M5S 1A7, Canada}
    \affiliation{Canadian Institute for Advanced Research, CIFAR Program in Quantum Materials, Toronto, Ontario M5G 1M1, Canada}
\date{\today}

\begin{abstract}
We study an extended Kitaev model on the triangular lattice in a limit where the symmetric off-diagonal bond-dependent and Heisenberg interactions together map onto an XXZ model, in addition to the Kitaev interaction. Within the previously identified $\mathbb{Z}_2$ vortex regime, we uncover a ferrichiral skyrmion phase characterized by a sublattice-resolved scalar chirality: two of the three sublattices carry unit skyrmion charge, while the third remains nonchiral. Using classical Monte Carlo simulations, we show that this ferrichiral skyrmion phase emerges at zero temperature and in the absence of both an external magnetic field and Dzyaloshinskii–Moriya interactions, in sharp contrast to conventional skyrmion-hosting systems. The phase is stable over a wide parameter window and persists to relatively high temperatures.
Our results reveal an unconventional route to skyrmion physics driven purely by frustrated exchange interactions and highlight the emergence of rich topological structures. Since both XXZ anisotropy and Kitaev interactions originate from the same spin–orbit–coupling mechanism, materials traditionally classified as XXZ magnets are expected to host finite Kitaev interactions as well. The potential for ferrichirality in these systems therefore warrants further investigation.
\end{abstract}

\maketitle

\section{Introduction}
Magnetic skyrmions are particle-like vortex spin textures characterized by an
integer-valued topological invariant, the skyrmion number.\cite{skyrmeNonlinearFieldTheory1961,yablonskiiThermodynamicallyStableVortices1989,bergDefinitionStatisticalDistributions1981}
Their nontrivial topology gives rise to emergent electrodynamic phenomena such as
the topological Hall effect and the skyrmion Hall effect.
\cite{zangDynamicsSkyrmionCrystals2011,schulzEmergentElectrodynamicsSkyrmions2012,nagaosaTopologicalPropertiesDynamics2013,iwasakiCurrentinducedSkyrmionDynamics2013,jiangDirectObservationSkyrmion2017}
Combined with the intrinsic robustness, these properties have positioned
skyrmions as promising information carriers for spintronic and nanodevice
applications.\cite{fertSkyrmionsTrack2013,zhangSkyrmionelectronicsWritingDeleting2020}

Magnetic skyrmions 
 were first experimentally realized in the chiral magnet
MnSi~\cite{muhlbauerSkyrmionLatticeChiral2009}, where they are stabilized by the
Dzyaloshinskii--Moriya (DM) interaction arising from the interplay of spin--orbit
coupling and inversion-symmetry breaking. This discovery triggered extensive
searches for skyrmion-hosting phases in non-centrosymmetric magnets, including
chiral systems such as MnSi and Fe$_{1-x}$Co$_x$Si
\cite{muhlbauerSkyrmionLatticeChiral2009,yuRealspaceObservationTwodimensional2010},
as well as polar magnets such as GaV$_4$X$_8$ (X = S, Se) and VOSe$_2$O$_5$
\cite{kezsmarkiNeeltypeSkyrmionLattice2015,kurumajiNeelTypeSkyrmionLattice2017a}.
More recently, it has become clear that inversion symmetry breaking is not a
prerequisite for skyrmion formation. Skyrmion phases have been theoretically
and experimentally identified in centrosymmetric systems stabilized by
geometric frustration \cite{okuboMultiple$q$StatesSkyrmion2012,kurumajiSkyrmionLatticeGiant2019},
RKKY interactions \cite{paddisonMagneticInteractionsCentrosymmetric2022a},
and competing orbital-dependent exchange processes
\cite{nomotoFormationMechanismHelical2020a,nomotoInitioExplorationShortpitch2023}.
Despite these advances, the role of frustrated exchange interactions alone in
generating skyrmion textures, particularly in the absence of magnetic fields and
DM interactions, remains relatively unexplored.

In parallel, there has been intense interest in frustrated spin systems where
bond-dependent exchange interactions emerge from the combined effects of strong
spin--orbit coupling, crystal-field environments, and Hund’s coupling.
A paradigmatic example is the Kitaev interaction, originally proposed as a route
to quantum spin liquids \cite{Kitaev2006}, and now realized in an expanding class
of materials \cite{Jackeli2009,Krempa2014SOC,Rau2015SpinOrbit,Hermanns2018Kitaev,Winter2017Kitaev,Perkins2024ROPP,matsuda2025rmp}.
Beyond the honeycomb lattice, the Kitaev interaction on the triangular lattice
has attracted particular attention.\cite{Jackeli2015TriangularKitaev, shinjoDensityMatrixRenormalizationGroup2016,Li2015_Triangular_Kitaev,Kos2017PRB,Maksimov2019_Triangular_AnisotropicExchange,Morita2020PRR,Wang2021PRB,Bhattacharyya2023NaRuO2,Razpopov2023_NaRuO2_Triangular_Kitaev,Xie2024PRL} Previous studies of the
Heisenberg--Kitaev model on the triangular lattice reported a broad parameter
regime hosting an unconventional magnetic state, commonly referred to as the
$\mathbb{Z}_2$ vortex phase, characterized by a nontrivial vortex pattern
\cite{beckerSpinorbitPhysics$jfrac12$2015,rousochatzakisKitaevAnisotropyInduces2016}.

Motivated by these findings, we study the triangular-lattice Kitaev system and
demonstrate that frustrated bond-dependent interactions alone are sufficient to
generate nontrivial topological textures. Specifically, we focus on an extended
Kitaev model on the triangular lattice including Heisenberg ($J$), Kitaev ($K$) and off-diagonal
$\Gamma$ and $\Gamma'$ interactions in the $\Gamma=\Gamma'$ limit, which maps to XXZ + Kitaev model as we will show below.
We uncover a finite skyrmion number hidden within the previously identified
$\mathbb{Z}_2$ vortex phase, manifested through a sublattice-resolved scalar
chirality, where two out of three sublattices carry a unit skyrmion charge. Our results establish that a topologically nontrivial {\it ferrichiral}
state emerges at zero magnetic field and in the absence of DM interactions,
revealing a new exchange-frustration-driven mechanism for skyrmion formation.

The paper is organized as follows.
In Sec. II, we review the bond-dependent interaction in a triangular lattice. We show that when $\Gamma = \Gamma'$, the model maps to XXZ + Kitaev interactions. In Sec. III, using classical Monte Carlo (MC) simulations, we map out the phase diagram of the XXZ + Kitaev model in the antiferromagnetic (AFM) regime and identify a broad vortex phase. While this phase was previously interpreted as a zero-temperature analog of thermally excited $\mathbb{Z}_2$ vortices in the AFM Heisenberg model \cite{kawamuraPhaseTransitionTwoDimensional1984}, we show that it also hosts
a nontrivial topological structure.
In Sec. IV, we define
 the skyrmion number separately
on the three sublattices, and show that a single vortex carries a total charge
$Q=\pm2$ from two sublattice leading to $\pm1$, while the third one gives no contribution. This ferrichiral texture can be understood as a superposition of two
Bloch-type skyrmions with opposite helicities residing on two of the three
sublattices, while the remaining sublattice forms a nonchiral background.
We summarize our findings and discuss possible candidate materials in the last section. 

\section{Review: Bond-dependent Hamiltonian and Z2 vortex phase}
In spin-orbit coupled Mott insulators with edge-sharing octahedra, magnetic exchange interactions acquire strong anisotropies beyond the conventional Heisenberg form. Jackeli and Khaliullin first showed that in the strong spin--orbit coupling limit, virtual superexchange processes generate a bond-dependent Ising interaction referred to as the Kitaev coupling $K$ on the honeycomb lattice, in addition to a Heisenberg term $J$~\cite{Jackeli2009}. Building on this framework, Rau, Lee, and Kee derived the most general nearest-neighbor spin Hamiltonian allowed by symmetry and explicitly introduced the symmetric off-diagonal exchange interactions $\Gamma$ and $\Gamma'$~\cite{Rau2014_Gamma}. The resulting $J$-$K$-$\Gamma$-$\Gamma'$ model has since become a minimal description of spin-orbit coupled honeycomb magnets such as the iridates and $\alpha$-RuCl$_3$.\cite{Plumb2014PRB,Kim2015KitaevMagnets,Sears2015MagneticOrder,Sandilands2015RamanContinuum,Johnson2015MonoclinicStructure,KimKee2016RuCl3AbInitio,Banerjee2016ProximateKitaev} More recently, analogous bond-dependent interactions have been shown to arise on the triangular lattice, where the same spin-orbit entangled superexchange mechanisms lead to generalized $J$-$K$-$\Gamma$ models and stabilize a variety of noncollinear and chiral magnetic states.\cite{Li2015_Triangular_Kitaev,catuneanuMagneticOrdersProximal2015,Maksimov2019_Triangular_AnisotropicExchange,Wang2021PRB,Razpopov2023_NaRuO2_Triangular_Kitaev}

\begin{figure}[!ht]
    \centering
    \includegraphics[width=\columnwidth]{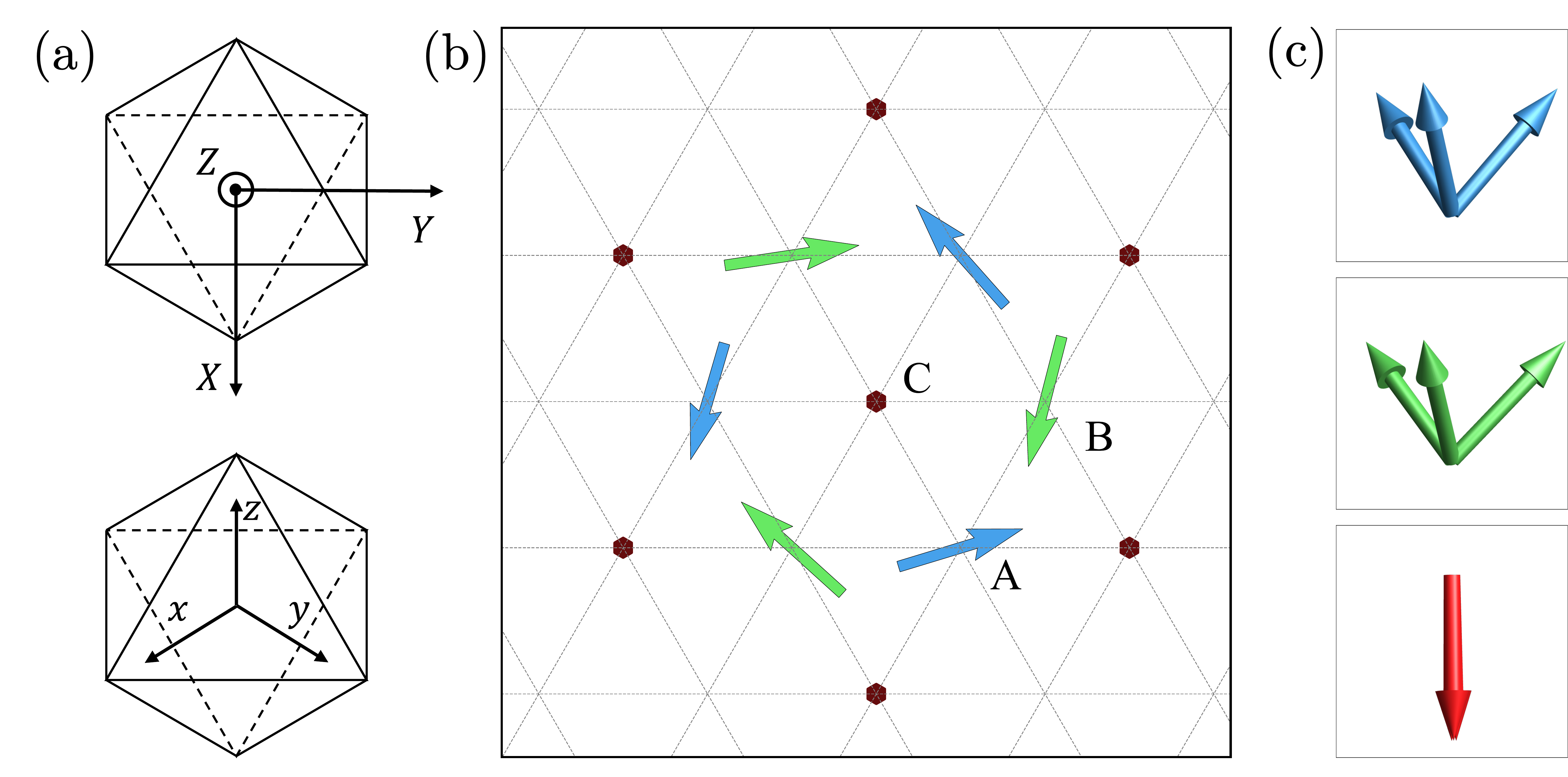}
    \caption{(a) Global coordinates $(X,Y,Z)$ and local octahedral coordinates $(x,y,z)$. (b) Spins at the center of the vortex from Fig.\ref{fig:phase_diagram}(f), colored according to their sublattice belonging, A-sublattice (blue), B-sublattice (green), and C-sublattice (red). (c) 3D plot of spins in different sublattices. It can be found that in A-sublattice (blue) and B-sublattice (green), spins are forming vortex, and they both contribute a positive value to the sublattice-dependent skyrmion number, whereas spins in C-sublattice (red) are forming non-chiral background.}
    \label{fig:coor+VC_center}
\end{figure}

The $J-K-\Gamma-\Gamma'$ model given as follows\cite{Rau2014_Gamma,chaloupkaHiddenSymmetriesExtended2015}:
\begin{equation}\label{eq:JKGG}
    \begin{aligned}
        H = &J\sum_{\braket{ij}}\vec{S}_i \cdot \vec{S}_j + K\sum_{\braket{ij}}S_i^\gamma S_j^\gamma + \Gamma \sum_{\braket{ij}} (S_i^\alpha S_j^\beta + S_i^\beta S_j^\alpha) \\
        & +\Gamma' \sum_{\braket{ij}} (S_i^\alpha S_j^\gamma + S_i^\gamma S_j^\alpha + S_i^\beta S_j^\gamma + S_i^\gamma S_j^\beta),
    \end{aligned}
\end{equation}
where $\gamma = x, y, z$ are the local octahedral coordinates. For a given $\gamma$, $\alpha$, $\beta$ are the other two coordinates as shown in Fig. \ref{fig:coor+VC_center}(a), and $\braket{ij}$ are nearest neighbor sites.

The Hamiltonian can be written in the crystallographic $(X,Y,Z)$ coordinate as shown in Fig. \ref{fig:coor+VC_center}(a), which offers experimentally relevant insights:
\begin{equation}
\renewcommand{\arraystretch}{1.5}
\begin{array}{l}
{\cal H}_{ij,\gamma}\!=\!J_{ab}(S_{i}^{X}S_{j}^{X}+S_{i}^{Y}S_{j}^{Y})+J_{c}S_{i}^{Z}S_{j}^{Z}\\~~
+A\left[c_\gamma(S_{i}^{X}S_{j}^{X}-S_{i}^{Y}S_{j}^{Y})-s_\gamma(S_{i}^{X}S_{j}^{Y}+S_{i}^{Y}S_{j}^{X})\right]\\~~
-\sqrt{2}B\left[c_\gamma(S_{i}^{X}S_{j}^{Z}+S_{i}^{Z}S_{j}^{X})+s_\gamma(S_{i}^{Y}S_{j}^{Z}+S_{i}^{Z}S_{j}^{Y})\right],
\end{array}
\label{eq:Hamiltonian2}
\end{equation}
where $c_\gamma\!\equiv\!\cos\phi_\gamma$, $s_\gamma\!\equiv\!\sin\phi_\gamma$, $\phi_{\gamma}\!=\!0$, $\frac{2\pi}{3}$, and $\frac{4\pi}{3}$ for $\gamma\!=\!z$-, $x$-, and $y$-bond, respectively, and 
\begin{equation}
    \label{JabJcABcouplings}
\begin{split}
A=&\frac{1}{3}K+\frac{2}{3}(\Gamma - \Gamma'), \;\; B=\frac{1}{3}K-\frac{1}{3} (\Gamma - \Gamma'),\\
J_{ab}&=J+B - \Gamma',\;\; J_{c}=J+A + 2 \Gamma'\,.
\end{split}
\end{equation}

When setting $\Gamma = \Gamma'$, the last two terms in Eq.\eqref{eq:JKGG} adds up to a bond-independent term;
$S_i^\alpha + S_i^\beta + S_i^\gamma = \sqrt{3} S_i^Z$, where $Z$ is one of the crystal axes, and the two terms of $\Gamma$ and $\Gamma^\prime$ in the Hamiltonian becomes:
\begin{equation}
    \begin{aligned}
        &\phantom{=} S_i^\alpha S_j^\beta + S_i^\beta S_j^\alpha + S_i^\alpha S_j^\gamma + S_i^\gamma S_j^\alpha + S_i^\beta S_j^\gamma + S_i^\gamma S_j^\beta \\
        &= (S_i^\alpha + S_i^\beta + S_i^\gamma)\cdot(S_j^\alpha + S_j^\beta + S_j^\gamma) - \vec{S}_i \cdot \vec{S}_j.
    \end{aligned}
\end{equation}
Then the Hamiltonian reduces to an anisotropic XXZ model in addition to the Kitaev interaction.
\begin{equation}
\renewcommand{\arraystretch}{1.5}
\begin{array}{l}
{\cal H}_{ij,\gamma}\!=\! J_{XY}(S_{i}^{X}S_{j}^{X}+S_{i}^{Y}S_{j}^{Y})+J_{Z} S_{i}^{Z}S_{j}^{Z}\\~~
+\frac{K}{3} \left\{ {\bf S}_i \cdot {\bf S}_j + c_\gamma(S_{i}^{X}S_{j}^{X}-S_{i}^{Y}S_{j}^{Y})-s_\gamma(S_{i}^{X}S_{j}^{Y}+S_{i}^{Y}S_{j}^{X})\right.\\~~
-\sqrt{2}\left.\left[ c_\gamma(S_{i}^{X}S_{j}^{Z}+S_{i}^{Z}S_{j}^{X})+s_\gamma(S_{i}^{Y}S_{j}^{Z}+S_{i}^{Z}S_{j}^{Y})\right] \right\},
\end{array}
\label{eq:Hamiltonian3}
\end{equation}
where $J_{XY} = J -\Gamma$, $J_{Z} = J + 2\Gamma$. The Kitaev interaction in the $XYZ$ coordinate contains the Heisenberg and bond-dependent parts. 
We denote this model as the XXZ + Kitaev model, and set $K \equiv J_0 \sin{\phi}$, $J_Z = J_0 \cos\phi$, and $J_{XY} = \epsilon J_0 \cos\phi$ to investigate the phase diagram as functions of $\epsilon$ and $\phi$. 

\begin{figure*}[!ht]
    \centering
    \includegraphics[width=0.95\textwidth]{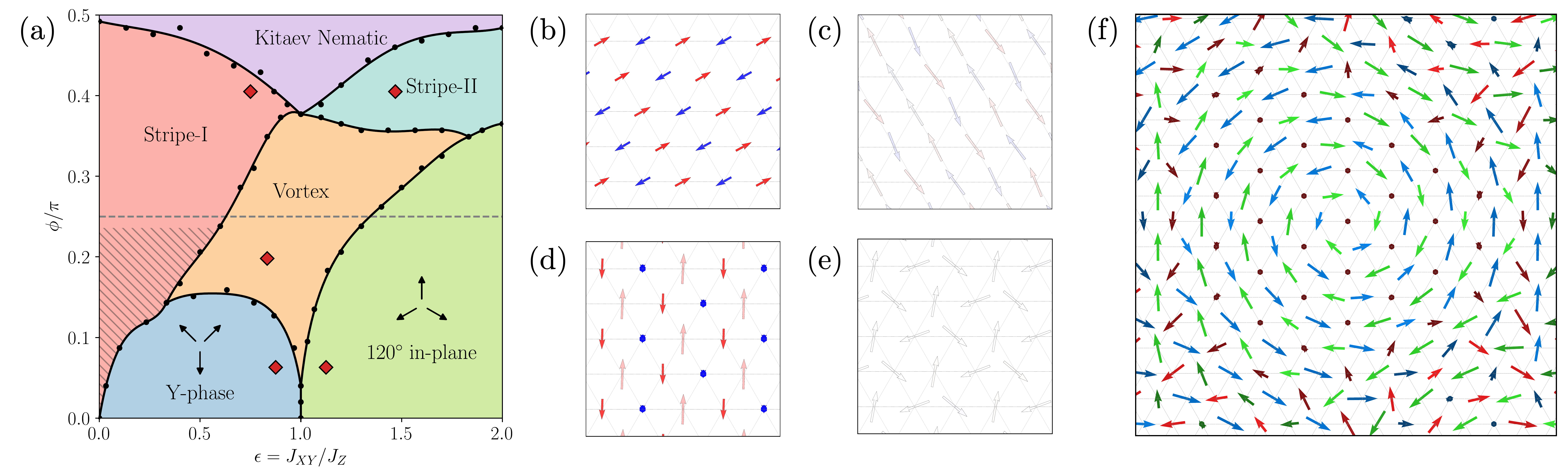}
    \caption{(a) Classic phase diagram obtained via Monte Carlo down to $T/J_0 = 10^{-6}$ in a $36 \times 36$ sites triangular lattice. The solid lines between phases represent second order phase transitions. The dashed area in Stripe-I phase represent a frustrated region that is unstable in Monte Carlo simulation and requires a future study. The dashed line corresponds to the line parameters used in Fig.\ref{fig:phi_cut}. (b)-(e) Real space spin configuration of the four ordered phases surrounding the vortex phase, calculated at parameters indicated by the diamonds in (a). They are: (b) the Stripe-I phase, (c) the Stripe-II phase, (d) the Y-phase, (e) the $120\degree$ in-plane phase. The arrow represents the projections of spins in $xy$-plane, whereas the color represents the $S_z$ components. (f) Real space spin configuration of the vortex phase, calculated at the diamond point in (a), where $\phi/\pi = 0.198$, $\epsilon=0.833$. Different colors represents different sublattices; whereas the brightness corresponds to the $z$-component of spins. In this case, the spins in the two sublatticies forming vortices (colored with blue and green) have mostly positive $S_z$, whereas in the other sublattice spins have mostly negative $S_z$.}
    \label{fig:phase_diagram}
\end{figure*}

 Before we present the phase diagram of the model, let us discuss a special limit. When $\epsilon = 1$, the model becomes the Heisnberg-Kitaev model, whose phase diagram was studied extensively in both classic limit and the quantum one.
In the classic limit, starting from the AFM Heisenberg point, when turning on the Kitaev interaction, a vortex phase immediately emerges, and connects all the way to the pure AFM Kitaev limit\cite{rousochatzakisKitaevAnisotropyInduces2016,kishimotoGroundStatePhase2018}.
Whereas in the quantum limit, numerous possibilities of this intermediate phase were suggested, including chiral spin liquid and $\mathbb{Z}_2$ vortex.\cite{liGlobalPhaseDiagram2015,kishimotoGroundStatePhase2018}
This vortex phase has been understood as an analog to the $\mathbb{Z}_2$ vortices in the anti-ferromagnetic Heisenberg model at finite temperature, which are considered as the nontrivial topological vortices or singular points defects in the $SO(3)$ order parameter field, corresponding to its first homotopy group: $\pi_1(SO(3)) = \mathbb{Z}_2$\cite{kawamuraPhaseTransitionTwoDimensional1984,rousochatzakisKitaevAnisotropyInduces2016}.
Similar vortices also exist in XXZ model at finite temperature, where instead they are $\mathbb{Z}$ vortices, since $\pi_1(U(1)) = \mathbb{Z}$.
By moving $\epsilon$ away from $1$, the XXZ anisotropic is turned on, and the Heisenberg-Kitaev model is extended to a XXZ-Kitaev model. This parametrization allows us to study the effect of both the bond-dependent Kitaev interaction and XXZ type anisotropy.

\section{classical phase diagram of the XXZ + Kitaev model}
To obtain the global phase diagram, we employ the classic MC method. 
The simulations are performed on a triangular lattice of 36 $\times$ 36 sites with periodic boundary conditions. A simulated annealing algorithm was used to cool the system down to $T/J_0 = 10^{-6}$, with in total $70,000$ MC steps.
We parametrize the coefficients as: $K=J_0\sin\phi$, $J_{Z} = J_0 \cos\phi$ and $J_{XY} = \epsilon J_0\cos\phi$, where $J_0$ set the energy scale, $\epsilon$ characterizes the XXZ anisotropy, and $\phi$ tunes the relative strength between the Kitaev and XXZ interactions.
We will set $J_0=1$ as the energy unit in the following discussion. To study the effect of both the bond-dependent Kitaev interaction and XXZ type anisotropy, we focus on the region close to the AFM Heisenberg axis, for which we chose $\phi \in [0, \pi/2]$ and $\epsilon\in[0.5,1.5]$.
The results were summarized in Fig.~\ref{fig:phase_diagram}.

When $\phi$ is set to $0$, the model becomes the XXZ model, whose AFM region was known to be dominated by two coplanar phases: the Y-phase in easy-axis region ($\epsilon<1$) and the $120\degree$ in-plane phase in easy-plane region ($\epsilon>1$), as demonstrated in Fig.~\ref{fig:phase_diagram}(d) and (e), respectively. 
Two phases are separated by the quantum critical point at $\epsilon=1$, where the AFM Heisenberg model exists and the ground state is identified as the $120\degree$ phase.\cite{kawamuraPhaseTransitionTwoDimensional1984}
When moving away from the $\phi=0$, both the Y-phase and the $120\degree$ in-plane phase extend to a broad area of our phase space. The Y-phase survives up to $\phi=0.15\pi$, whereas the $120\degree$ in-plane phase survives up to $\phi=0.25\pi$, before the bond anisotropic Kitaev interaction starts to dominate.

In the classical pure Kitaev limit ($\phi = \pi/2$), the ground state is a nematic phase characterized by the system decomposing into decoupled antiferromagnetic Ising chains, where spins are Ising-aligned along their bond-dependent easy axes.
The decoupling of these chains leads to a subextensive ground-state degeneracy of $3 \times 2^{\sqrt{N}}$, though it would be reduced to $3 \times 2^2$ when quantum fluctuation exists.\cite{jackeliQuantumOrderDisorder2015}
This nematic phase is an extended phase when turning on the bond-independent XXZ interaction, but would eventually undergo phase transition into two stripe phases, labeled as Stripe-I and Stripe-II in Fig.~\ref{fig:phase_diagram}(b) and (c), respectively.
The Stripe-I phase is a collinear phase with spins aligned to the Kitaev easy axes, whereas in the Stripe-II phase spins are aligned within the lattice plane, with respect to its easy-plane type XXZ interaction.

As the $\phi$ is turned into an intermediate regime, a vortex phase emerges in the center of the four ordered phases, as demonstrated in Fig.~\ref{fig:phase_diagram}(f).
This phase has been studied in the Heisenberg-Kitaev model ($\epsilon = 1$), where it was identified as the $\mathbb{Z}_2$ vortex phase\cite{rousochatzakisKitaevAnisotropyInduces2016,beckerSpinorbitPhysics$jfrac12$2015}.
It survives extensively from $\epsilon \approx 0.4$ to $\epsilon \approx 1.8$, indicating its robustness against competition with other ordered phases.

\begin{figure}[!h]
    \centering
    \includegraphics[width=0.8\columnwidth]{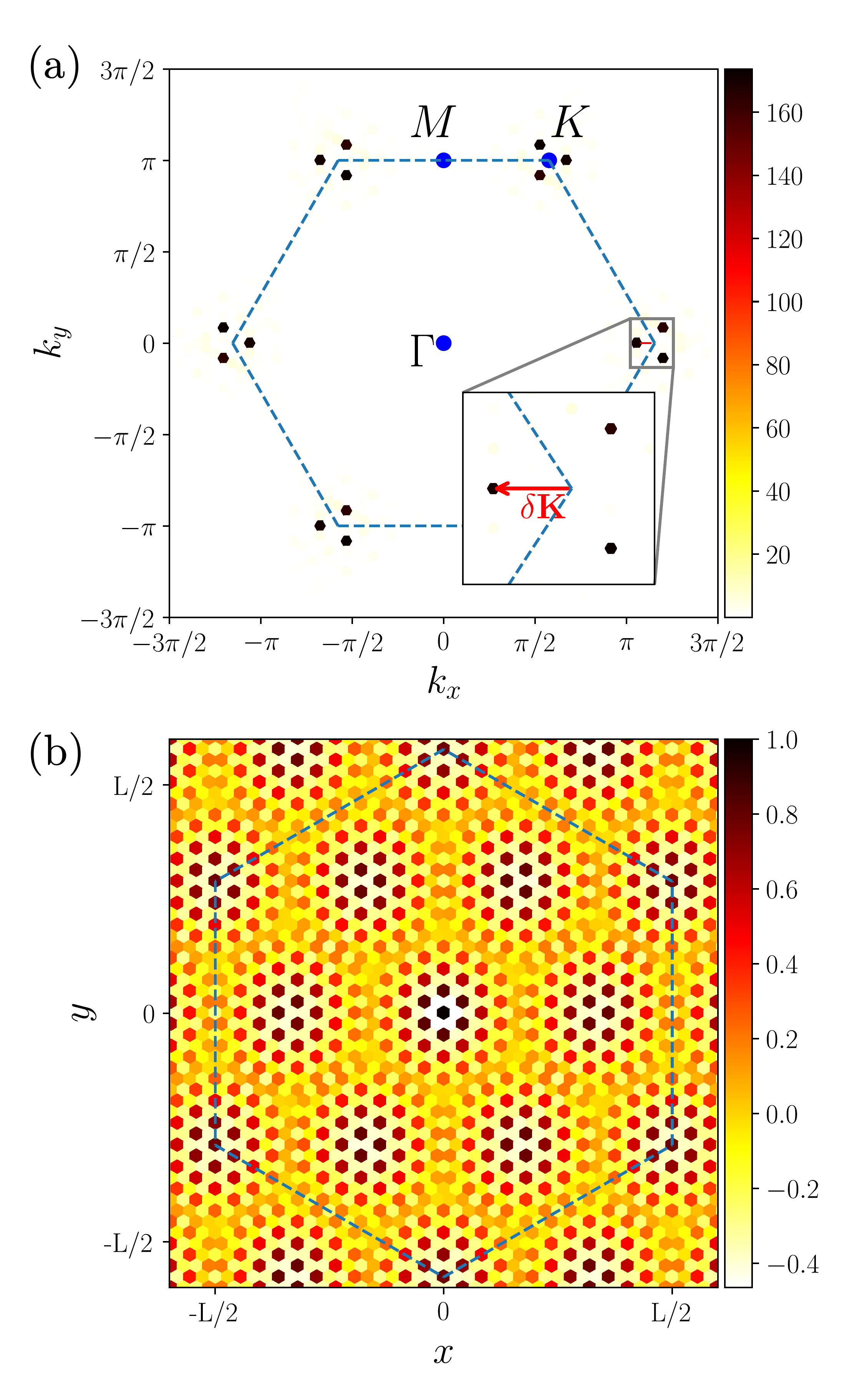}
    \caption{Static structure factors of the vortex phase. (a) The static structure factor $\braket{\vec{S}\cdot\vec{S}}(\mathbf{q})$ in momentum space, with triple-q satellite peaks (black dots) around the K points. The dashed line outlines the first Brillouin zone where the special momentum points such as $K$, $M$, and $\Gamma$ are denoted by the blue dots. The inset shows the zoom-in area near the K-point to indicate the $\delta{\bf K}$ characterizing the vortex density $n_{\rm vortices}$ defined in Eq. (6). (b) The real-space spin correlation $\braket{\vec{S}_{\mathbf{R}+\mathbf{r}}\cdot\vec{S}_{\mathbf{R}}}_{\mathbf{R}}$, revealing the vortex superlattice. The dashed line indicates the lattice boundary with periodic boundary conditions.}
    \label{fig:Scorr_Z2VC}
\end{figure}

The vortex phase is characterized by the formation of a vortex superlattice. This is evidenced by the emergence of triple-q satellite peaks in the static spin structure factor, located around the K-points of the Brillouin zone, as shown in Fig.~\ref{fig:Scorr_Z2VC}. The vortex density, $n_{\text{vortices}}$, defined the number of vortices per site, can be determined from the deviation of these satellite peaks, ${\bf Q}_{\text{peak}}$, from the K-points as follows.
\begin{equation}
    n_{\text{vortices}} = \frac{N_{\text{vortices}}}{N_{\text{sites}}} = \frac{|\delta {\bf K}|^2}{|{\bf K}|^2},
    \label{eq:vortex_density}
\end{equation}
where $\delta {\bf K} = {\bf K}-{\bf Q}_{\text{peak}}$, as denoted by the red arrow in the inset of Fig.\ref{fig:Scorr_Z2VC}(a).

The vortex density, $n_{\text{vortices}}$, characterized by the deviation $|\delta \mathbf{K}|$ in the static structure factor, can vary within the vortex phase.
As $\phi$ increases, $Q_{\text{peak}}$ shifts from the K-point towards the midpoint between two neighboring M-points, and $|\delta \mathbf{K}|$ increases concurrently from $0$ to $|\mathbf{K}|/4$.
When $|\delta \mathbf{K}|$ reaches its maximum, the satellite peaks $Q_{\text{peak}}$ will then either broaden into arcs connecting neighboring M-points or jump directly onto the M-points, meanwhile spontaneously breaking the $C_3$ symmetry and selecting a single direction. 
The former process will lead the system transitioning into the Kitaev nematic phase, and the latter will lead to one of the two stripe phases.

In finite-size MC, strong finite size effects may hinder in determining
 $|\delta {\bf K}|$ for a given phase space. Since the Brillouin zone of a finite lattice permits only discrete $k$-points, $|\delta {\bf K}|$ will change discontinuously, resulting in behaviors similar to a phase transition. 
In the thermodynamic limit, however, we expect that the vortex density will continuously vary inside the vortex phase, as indicated by the previous Luttinger-Tizsa result\cite{rousochatzakisKitaevAnisotropyInduces2016}, and thus the discontinuity is therefore an artifact of the finite size MC cluster.

The ground state in the shaded area around the left bottom of the phase diagram close to the $\epsilon=0$ line is not clearly identified, as the MC calculations do not fully converge to a stable configuration. This might be subjected to the strong frustration near the Ising point $\epsilon=0$, $\phi=0$, which warrants further study in future work.

\section{Sublattice-Resolved Skyrmion}
This vortex phase, extending continuously from the Heisenberg-Kitaev model, was previously identified as the $\mathbb{Z}_2$ vortex phase, in analog to that in pure AFM Heisenberg model at finite temperature\cite{rousochatzakisKitaevAnisotropyInduces2016,beckerSpinorbitPhysics$jfrac12$2015}.
However, two significant differences can be observed between the vortices in the XXZ+Kitaev model and the original $\mathbb{Z}_2$ vortices in the Heisenberg model.
First, when tuning on the Kitaev interaction,  neither $SO(3)$ nor $U(1)$ symmetry is present, reducing to the discrete symmetry of $D_{3d}$, on which homotopy groups are not well-defined.
Second, this phase forms at zero-temperature, which is much more stable than the finite-temperature case, and exhibits stable super-lattice structures.
These differences suggest that the vortex phase in the XXZ + Kitaev model may display richer properties.

As illustrated in Fig.~\ref{fig:coor+VC_center} (b) and Fig.~\ref{fig:phase_diagram}(f), each vortex can be decomposed into three sublattices with relatively simpler structures.
Inspired by this observation, we propose a sublattice-dependent skyrmion number as a characterization of the vortex phase in the XXZ + Kitaev model. Consider the system having a $\sqrt{3}\times\sqrt{3}$ magnetic unit cell, which divides the triangular lattice into three sublattices, labeled as $A$, $B$, and $C$. The skyrmion number on each sublattice is defined as:
\begin{equation}
    Q_{s} = \frac{1}{4\pi} \int S_s \cdot (\partial_x S_s \times \partial_y S_s) \md x \md y \equiv \frac{1}{4\pi} \int q_s(x,y) \md x \md y
\end{equation}
where $s=A,B,C$ labels the sublattice. Since it depends on the sublattice, we refer "sublattice-resolved" skyrmion number. In a discrete lattice, the integral is replaced with a sum over all triangle plaquettes $p$ in the sublattice. On each plaquette with vertices $i,j,k$, the integrand $q_s(x,y)$ can be reformed as follows\cite{bergDefinitionStatisticalDistributions1981}:
\begin{equation}
    q_{s,\braket{ijk}} = 2 \arctan \frac{\vec{S}_{i}\cdot(\vec{S}_{j}\times\vec{S}_{k})}{1+\vec{S}_{i}\cdot\vec{S}_{j}+\vec{S}_{j}\cdot\vec{S}_{k}+\vec{S}_{k}\cdot\vec{S}_{i}},
\end{equation}
which is exactly the solid angle formed by the three spin vectors $\vec{S}_{i}$, $\vec{S}_{j}$, and $\vec{S}_{k}$ on a unit sphere. The skyrmion number is then defined as the summation of $Q_s$ 
on the three sublattices:
\begin{equation}
    Q = \sum_{s=A,B,C} Q_{s} = \sum_{s} \sum_{\braket{ijk} \in s} q_{s,\braket{ijk}}
    \label{eq:mod_skyrmion}
\end{equation}

\begin{figure}[!t]
    \centering
    \includegraphics[width=0.9\columnwidth]{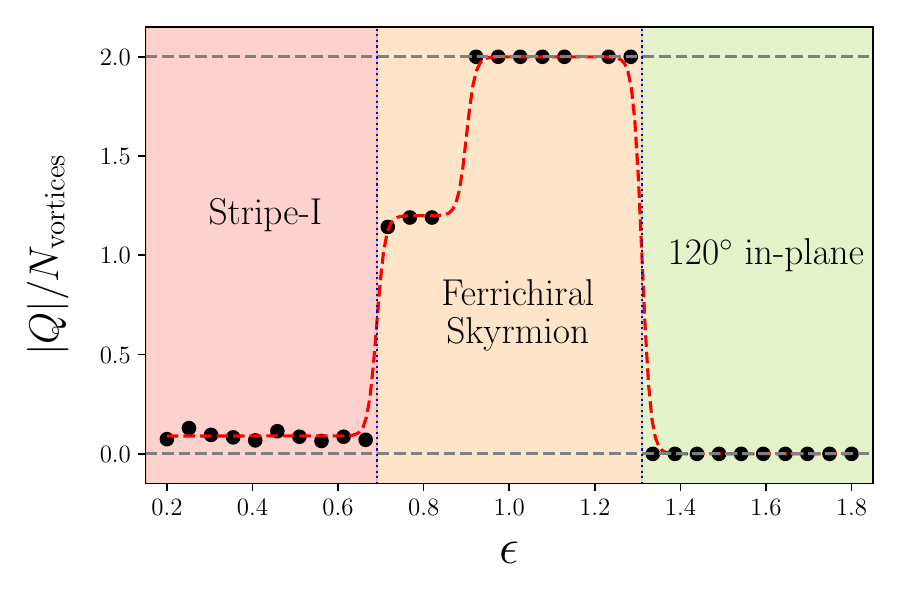}
    \caption{The sublattice-resolved skyrmion number per vortex, $|Q|/N_{\text{vortices}}$, computed along the dashed line in Fig.~\ref{fig:phase_diagram}(a) with $\phi=0.25\pi$ and $\epsilon \in (0.2,1.8)$, using MC method is shown as the black dots. The dashed red line is a guide to the eye. Two sharp transitions from regions where $|Q|/N_{\text{vortices}} \sim 0$ to a non-zero region occur at $\epsilon = 0.69$ and $\epsilon = 1.31$, in agreement with the phase boundaries in Fig.~\ref{fig:phase_diagram}(a). For $\epsilon < 0.69$, small finite values $\sim0.09$ were found, which is attributed to the adjacent unstable frustrated region. A narrow plateau at $|Q|/N_{\text{vortices}} \approx 1.2$ emerges within $\epsilon \in (0.69, 0.90)$, indicating that the vortices are not yet fully ordered in this region. Within $\epsilon\in(0.90, 1.31)$, $|Q|/N_{\text{vortices}}$ saturates at $2$, indicating that the vortices are aligned ferromagnetically in the entire lattice.}
    \label{fig:phi_cut}
\end{figure}

To reveal the non-trivial topological property of the vortex phase, we compute the sublattice-resolved skyrmion number per vortex $|Q|/N_{\text{vortices}}$ defined in Eq.~\eqref{eq:mod_skyrmion}.
The calculation is performed along a specific path in the phase diagram, as indicated in Fig.~\ref{fig:phase_diagram}(a), for a $36\times36$ lattice.
We employ the parallel tempering MC method across a temperature range $T/J_0 \in [0.01, 1]$. 
For each parameter point, the reported value is the maximum obtained within the low-temperature region ($T/J_0 < 0.04$), ensuring ground-state dominance.

Along this cut, we find $|Q|/N_{\text{vortices}} \approx 0$ within both the Stripe-I and the $120^\circ$ in-plane ordered phases.
In contrast, a finite value emerges inside the vortex phase, featuring two distinct plateaus.
The first plateau attains the maximal value of $2$, corresponding to a state where the cores of all vortices are uniformly aligned.
The second plateau resides at $|Q|/N_{\text{vortices}} \sim 1.2$, where approximately $4/5$ of the vortices share a common core orientation while the remaining $1/5$ are oppositely oriented.

This quantized behavior demonstrates that each individual vortex of sublattice A and B contributes $\pm 1$ to the total skyrmion number $Q$, which can be understood from its internal structure.
As illustrated in Fig.~\ref{fig:coor+VC_center}(b) and Fig.~\ref{fig:phase_diagram}(f), each vortex comprises three distinct sublattices A, B, and C.
Sublattices A and B host vortex-like spin textures identifiable as Bloch-type skyrmions. 
Since their helicities and vorticities are mutually opposite, their contributions to the sublattice-resolved skyrmion number $Q_s$ are identical: both $+1$ or both $-1$.
Meanwhile, sublattice C exhibits a nearly ferromagnetic alignment along the $Z$-axis in the vortex core, deviating only gradually at the periphery of the vortex, resulting in a net contribution of $0$ to $Q_s$.
Consequently, the net contribution from one vortex sums to $\pm 2$.
Since the polarization of C sublattice in the central region is anti-correlated with the $S_z$ component of sublattices A and B due to AFM couplings, the core polarization is thereby binded to the sign of the vortex's topological charge $Q$.

The quantization of $Q/N_{\text{vortices}}$ to $\pm 2$ reveals an emergent binary, or $\mathbb{Z}_2$, degree of freedom.
As in Fig.~\ref{fig:phi_cut}, the two plateaus with non-zeros $Q/N_{\text{vortices}}$ between $\epsilon\in(0.69, 1.31)$ suggest a tendency for this degree of freedom to develop long-range order.
To investigate its thermodynamic behavior, we perform large-scale parallel tempering simulations for a specific parameter point ($\phi = 0.25\pi$, $\epsilon=1.1$) on a $64 \times 64$ lattice.
We sample the temperature range $T/J_0 \in [0.02, 0.5]$ with $2\times10^6$ Monte Carlo steps per temperature.

\begin{figure}[!ht]
    \centering
    \includegraphics[width=\columnwidth]{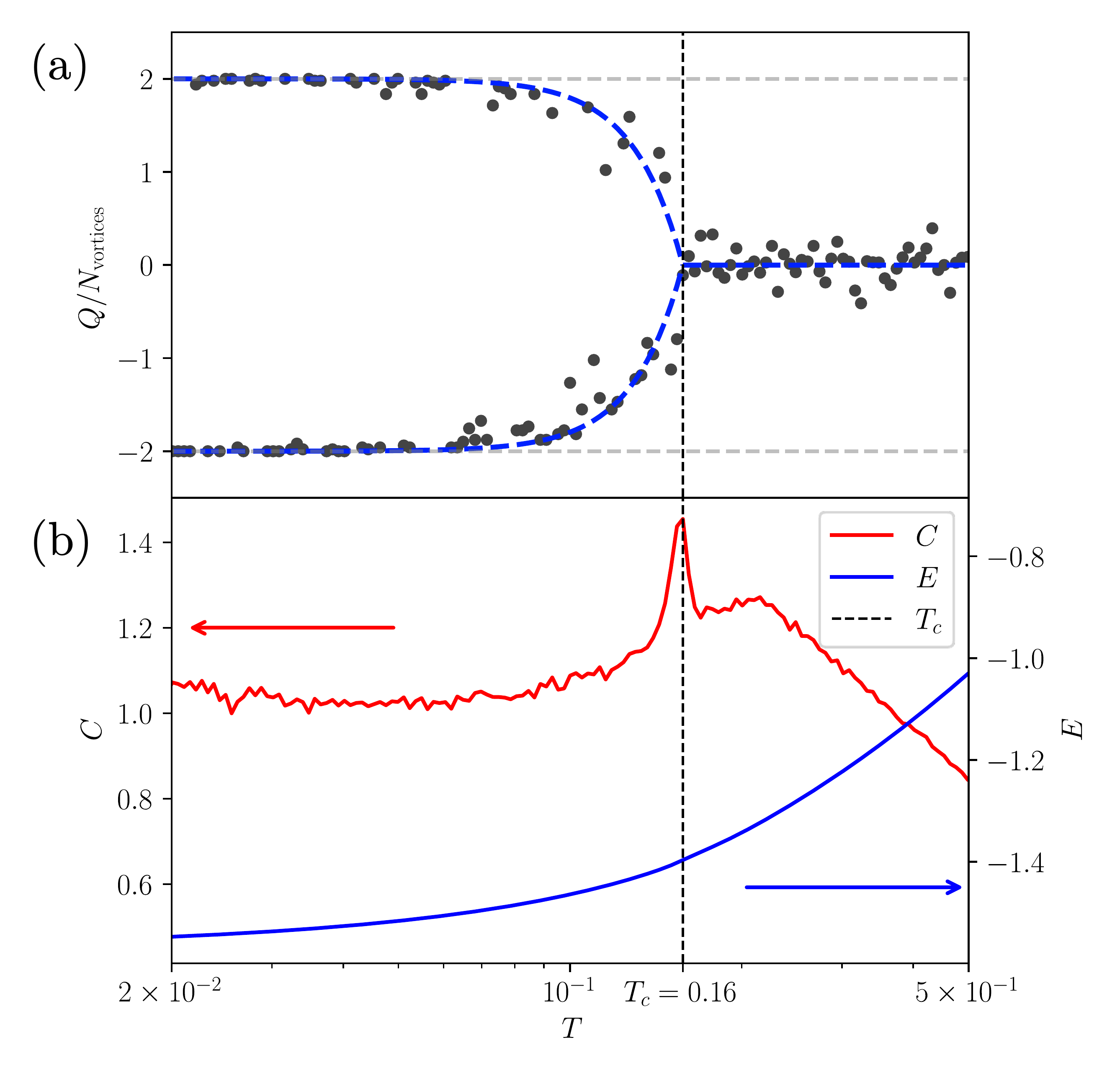}
    \caption{Calculation using parallel tempering on a $64\times64$ lattice, with parameters set to be $\phi = 0.25\pi$, $\epsilon=1.1$, and temperatures $T$ ranging from $0.02\sim 0.5$, in unit of $J_0$. $2\times10^6$ sweeps were done at each single temperature point. (a) $Q/N_{\text{vortices}}$ as a function of temperature $T$ obtained by Monte Carlo calculations is denoted by black dots. The blue dashed line is a guide to the eye. The vertical dotted line indicats the transition temperature $T_c$. $Q/N_{\rm vortices}$ saturates to $\pm 2$ at low temperatures. (b) Energy $E$ (blue) and specific heat $C$ (red) as a function of $T$. $T_c = 0.16J_0$ can be determined by the maximum of specific heat, which also corresponds to the onset of $Q/N_{\text{vortices}}$ as shown in (a).}
    \label{fig:L=64}
\end{figure}

The results are shown in Fig.~\ref{fig:L=64}. The specific heat exhibits a sharp peak indicative of a second-order phase transition, from which we identify a critical temperature $T_c \approx 0.16J_0$.
Above $T_c$, $Q/N_{\text{vortices}}$ fluctuates around zero, corresponding to a disordered paramagnetic state.
Below $T_c$, $|Q|/N_{\text{vortices}}$ grows rapidly while the vortex structure starts to form simultaneously.
As the system cools down to $T/J_0 \lesssim 0.07$, $Q/N_{\text{vortices}}$ saturates near the extremal value of $\pm 2$, and all vortices form a well-established superlattice structure, with their cores aligned, consistent with the $36\times36$ ground-state results in Fig.~\ref{fig:phi_cut}.

These findings confirm that, at least near the studied parameter point, the $\mathbb{Z}_2$ degrees of freedom order ferromagnetically, corresponding to the $|Q|/N_{\text{vortices}}=2$ plateau.
Although our finite-temperature simulations in other regions of the vortex phase do not always reach the saturated value of $\pm 2$, the observed $Q/N_{\text{vortices}}$ values are consistently large ($\sim \pm 1$).
Furthermore, we observe a strong tendency for neighboring vortices to share the same core orientation, suggesting prevalent ferromagnetic correlations throughout the vortex phase.
In light of this uniform $\mathbb{Z}_2$ order coupled with the chiral vortex texture, we term this phase the ferrichiral skyrmion state.

Finally, a finite-size scaling analysis across different lattice sizes yields an average critical temperature of $T_c \approx 0.19J_0$ for this parameter set.
This relatively high value ($T_c/J_0 \sim 0.2$) indicates that the effective interactions mediating the ferromagnetic alignment of vortex cores are not weak, underscoring the stability of the ferrichiral skyrmion phase.

\section{Summary and Discussion}
Since the identification of a microscopic mechanism generating bond-dependent Kitaev interactions on the two-dimensional honeycomb lattice \cite{Jackeli2009}, it has become clear that such interactions are ubiquitous in spin–orbit-coupled magnetic systems. When combined with lattice geometries beyond the honeycomb structure, such as the triangular lattice, the interplay between conventional Heisenberg exchange and Kitaev interactions has been shown to produce a rich phase diagram, including the emergence of a $\mathbb{Z}_2$ vortex phase.\cite{beckerSpinorbitPhysics$jfrac12$2015, catuneanuMagneticOrdersProximal2015,rousochatzakisKitaevAnisotropyInduces2016, shinjoDensityMatrixRenormalizationGroup2016}
Because the Kitaev interaction originates from strong spin–orbit coupling, additional symmetry-allowed bond-dependent interactions—commonly referred to as the $\Gamma$ and $\Gamma'$ terms—are generally unavoidable \cite{Rau2014_Gamma}. Notably, when $\Gamma = \Gamma'$, the resulting Hamiltonian can be viewed as an XXZ model supplemented by a bond-dependent Kitaev interaction. 

In this work, we study the XXZ + Kitaev model and demonstrate the emergence of a sublattice-dependent skyrmion structure within the $\mathbb{Z}_2$ vortex phase. Specifically, two of the sublattices carry a unit skyrmion charge, while the remaining sublattice hosts zero skyrmion charge. We therefore refer to this phase as a ferrichiral skyrmion state.
Our theory reveals an unconventional mechanism for stabilizing skyrmion phases, distinct from the conventional routes that rely on inversion-symmetry breaking and the resulting Dzyaloshinskii–Moriya interaction, or on explicit time-reversal-symmetry breaking either spontaneously broken or induced by external magnetic fields.
Notably, this phase emerges at relatively high temperatures, as demonstrated by classical Monte Carlo simulations.
The phase is also robust across a broad parameter range in the studied case ($\Gamma = \Gamma'$), hinting at its possible stability for $\Gamma \neq \Gamma'$.
These findings prompt further analysis to assess the phase's thermodynamic stability and the impact of deviations from the XXZ+Kitaev limit.

Our findings naturally motivate a discussion of material realizations. Since both XXZ anisotropy and Kitaev-type interactions arise from the same spin–orbit-coupling mechanism, materials previously described as XXZ magnets are expected to host finite Kitaev interactions as well. Consequently, triangular-lattice XXZ systems, such as the Yb-delafossites NaYbCh$_2$(Ch = O, S, Se)\cite{sichelschmidtElectronSpinResonance2019a,guoMagneticfieldCompositionTuned2020,daiSpinonFermiSurface2021} and their doped variant NaYb$_{1-x}$Lu$_x$Ch$_2$\cite{hausslerDilutingTriangularlatticeSpin2022,zhangMethodDetectingMagnetic2025,alvaradoMagneticDilutionTriangular2025}, Cs$_2$CuBr$_4$\cite{onoMagnetizationPlateauFrustrated2003,onoPhaseTransitionsDisorder2005} and Ba$_3$CoSb$_2$O$_9$\cite{sera$Sfrac12$TriangularlatticeAntiferromagnets2016, itoStructureMagneticExcitations2017, kamiyaNatureSpinExcitations2018} compounds, are promising candidates for realizing this physics.
Additionally, potential $Z_2$ vortex crystal signatures have been reported in materials such as NaCrO$_2$\cite{tomiyasuObservationTopological$mathbbZ_2$2022} and $\mathrm{(CD_3ND_3)_2NaRuCl_6}$\cite{nagl$mathbbZ_2$VortexCrystal2025a}. 
The roles of Kitaev and XXZ interactions in these materials warrant 
further theoretical and experimental investigation. 
It would be also worthwhile to explore their effects on transport properties such as anomalous Halll conductivity in future work.

\section{acknowledgment}
We thank Haoting Xu for useful discussion. This work is supported by the NSERC Discovery Grant No. 2022-04601 and NSERC CREATE program No. 575280-2023. H. Y. K. acknowledges support from the Canada Research Chairs Program No. CRC-2019-00147.
This research was enabled in part by support provided by Compute Ontario,  Calcul Québec, and the Digital Research Alliance of Canada.

\bibliographystyle{apsrev4-2}
\bibliography{ref} 

\end{document}